\documentclass[11pt, oneside]{article}   	

\usepackage{geometry}                		
\usepackage[parfill]{parskip}    		
\usepackage{graphicx}				
\usepackage{epsfig}
\usepackage{amsmath}
\usepackage{amssymb}
\usepackage{dsfont}
\usepackage{subfigure}
\usepackage[pagebackref=true,breaklinks=true,letterpaper=true,colorlinks,bookmarks=false]{hyperref}

\title{Building A Large Concept Bank for Representing Events in Video}
\author{Yin Cui, Dong Liu, Jiawei Chen, Shih-Fu Chang\\
Department of Electrical Engineering, Columbia University\\
}

\begin{document}
\maketitle

\begin{abstract}
\label{sec_abstract}
Concept-based video representation has proven to be effective in complex event detection.
However, existing methods either manually design concepts or directly adopt concept libraries not specifically designed for events.
In this paper, we propose to build Concept Bank, the largest concept library consisting of $4,876$ concepts specifically designed to cover $631$ real-world events. To construct the Concept Bank, we first gather a comprehensive event collection from WikiHow, a collaborative writing project that aims to build the world's largest manual for any possible How-To event.
For each event, we then search Flickr and discover relevant concepts from the tags of the returned images.
We train a Multiple Kernel Linear SVM for each discovered concept as a concept detector in Concept Bank.
We organize the concepts into a five-layer tree structure, in which the higher-level nodes correspond to the event categories while the leaf nodes are the event-specific concepts discovered for each event. Based on such tree ontology, we develop a semantic matching method to select relevant concepts for each textual event query,
and then apply the corresponding concept detectors to generate concept-based video representations. We use TRECVID Multimedia Event Detection 2013 and Columbia Consumer Video open source event definitions and videos as our test sets and show very promising results on two video event detection tasks: event modeling over concept space and zero-shot event retrieval.
To the best of our knowledge, this is the largest concept library covering the largest number of real-world events.
\end{abstract}

\section{Introduction}
\label{sec_introduction}
Representing and detecting complex events from unconstrained videos remains one of the most challenging problems in computer vision.
By definition, an event is a complex activity that involves people interacting with other people and/or objects under certain scene settings.
For instance, ``changing a vehicle tire'' can be defined as a complex event where human, vehicle, tire and tools interact with each other in an outdoor environment.
The existing research on video event detection focuses on the use of low-level features combined with sophisticated learning models, and achieves satisfactory performances to some extent~\cite{duan2009domain, natarajan2012fusion, tang2012latent}. However, these works fail to provide semantic information in a video event.
This hampers high-level event understanding, especially when the number of training videos is small or zero.
Therefore, a reasonable and computationally tractable way is to decompose videos depicting an event into a set of atomic concepts \cite{liu_concept, yang2012deepnet}.
These concepts can be perceived as building blocks of a complex event and are expected to provide a meaningful intermediate level of abstraction towards representing an event rather than low-level visual-audio features directly extracted from videos.
In this paper, we focus on such an approach in which the textual title of the event (\emph{e.g.}, ``grooming an animal") is used as query to first find concepts relevant to the event, and then we use the selected concepts as representations in supervised event modeling or zero-shot learning.

There are two existing approaches to generate concept based video representation.
The first is to manually define suitable concepts for each event, which involves too many human efforts and hence is not feasible for large-scale problems.
The second is to directly utilize the existing banks built on certain ontologies.
However, the main issue is that such banks are not specifically designed for complex events and do not contain enough relevant concepts.
Blindly applying these banks with significant amount of irrelevant concepts will degrade the very purpose of high-level concept-based video representation.

This motivates us to build an event-oriented concept library called ``Concept Bank'' (CB) for high-level representation of complex events in video.
In this paper, we identify three challenges for building such a concept library and make significant contributions in developing corresponding solutions summarized below.

The first challenge is that the library should be \textbf{quasi-comprehensive} and cover as many real-world events as possible.
To address this challenge, we collect and select events from Wikihow \cite{wikihow}, a collaborative writing project that aims to build the world's largest how-to manual for any possible event\footnote{Although WikiHow only focuses on how to do anything, we realize that it is created based on user's common interests and thus has good coverage on almost every aspect of human daily life.
Advanced event collection techniques can be also used to extract visually detectable events from other knowledge bases, and easily incorporated into our current system.}.
In this way, we end up with $631$ events as our event collection. To the best of our knowledge, this is the largest event collection in the literature.

The second is how to \textbf{discover concepts} and build their visual models for each event.
To solve this, we apply our recently proposed idea on automatic event-driven semantic concept discovery from Web images~\cite{FlickrConcept}.
The web is a rich source of information with tremendous images captured for various events and these images are roughly annotated with descriptive tags that indicate the semantics of the image contents.
Our intuition is that if we crawl enough number of images by an event query, tags of these images could somehow reveal certain semantics related to the event statistically, and suggest relevant concepts of the event.
In light of this, we crawl images and their associated tags from Flickr using event queries extracted from WikiHow, then find the most relevant event-specific concepts from the tags and build visual model for each concept.

Finally, after we get a comprehensive concept library for all events, it remains unclear on how to choose the most \textbf{relevant concepts} to represent any possible query event.
To address this task, we propose a semantic matching method to measure the relevance between each concept and the query event
(without using any training videos).
We then choose the top ranked concepts to represent the event and get a compact concept based video representation.

\begin{figure}[!ht]
\centering
\includegraphics[width=1\columnwidth]{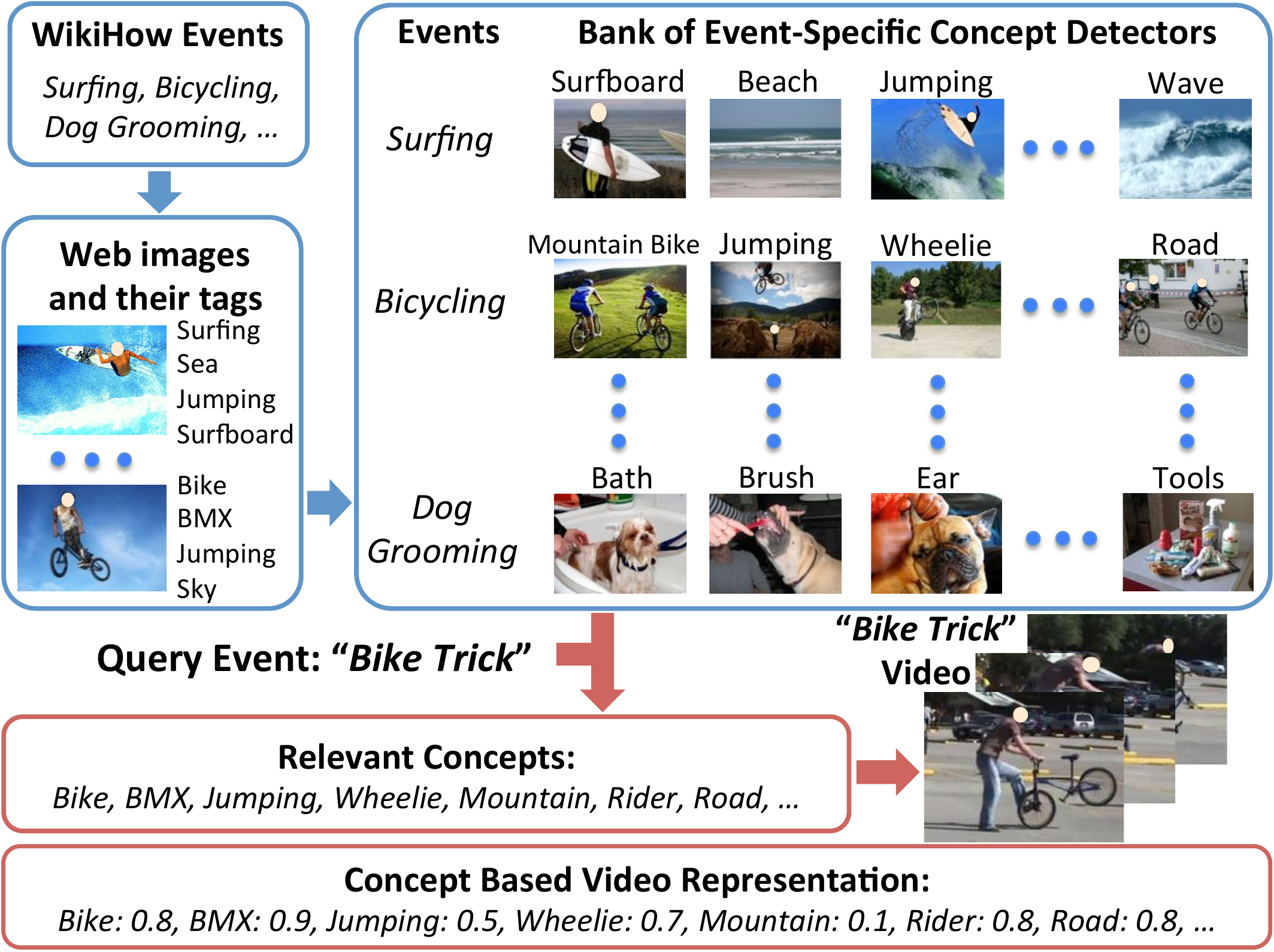}
\caption{\textbf{Building Concept Bank} (in the upper three boxes): We collect a comprehensive event list from WikiHow.
For each event, we search Flickr images and discover relevant concepts from their tags~\cite{FlickrConcept}.
We train a concept detector for each discovered concept, resulting a large set of event-specific concept detectors.
\textbf{Video Representation by Concept Bank} (in the lower part):
Given a query event, relevant concepts are selected based on semantic matching (Section~\ref{sec_representation}).
Then the selected concept detectors are applied on frames of a video clip, generating a concept score vector on each frame.
Finally, we use average-pooling to aggregate all concept score vectors across the frames.}
\label{fig_overview}
\end{figure}

Figure~\ref{fig_overview} illustrates an overview of how we build the proposed Concept Bank and use it to represent an event video.
Experimental results show that the video representation with only few hundred dimensions generated by CB achieve state-of-the-art event detection performance.

\newpage

\newpage
\section{Related Work}
\label{sec_related_work}
Video event detection has became an important research area in computer vision literature. 
Natarajan \emph{et al.} \cite{natarajan2012fusion} investigated the multimodal fusion of low-level video features and the extracted videotext information in video content, and achieved good performance on the detection. Ye \emph{et al.} \cite{bimodal} discovered the bi-modal audio-visual codewords and leveraged the joint patterns across the audio and visual space to boost event detection performance. Duan \emph{et al.} \cite{duan2009domain} incorporated the web sources videos crawled from Youtube to relieve the insufficiency of the number of training videos of an event, and developed a cross-domain video event detection model. Tang \emph{et al.} \cite{tang2012latent} developed a large margin framework to exploit the latent  temporal structure in successive clips of a long event video. These excellent works focus on modeling events into sophisticated statistical models or fusing mutimodal information. However, none of them can reveal the rich semantics in event videos.  

There are some recent works that try to perform event detection with semantic concepts. 
Yang \emph{et al.} \cite{yang2012deepnet} applied deep belief nets to group a large number of event video shots into a number of shot clusters, and then treat each cluster center as a data-driven concept. Then each video is mapped onto the cluster centers and encoded into a concept based representation.
However, such data-driven concepts do not convey any semantic information and hence cannot be utilized for high-level semantic understanding.
Liu \emph{et al.} \cite{liu_concept} manually defined concepts present in event videos and categorized them into ``object'', ``scene'' and ``action''.
They annotated the presence of each concept on training videos and then built the individual concept detectors.
Nevertheless, as aforementioned, this requires too many manual efforts, and is not applicable for real-world large-scale video event detection tasks.
Different from these prior works, we focus on automatically discovering potential concepts present in any possible events with interpretable semantic meaning.

Notably, we applied our previous work on concept discovery from Internet images~\cite{FlickrConcept} as a building component of our concept library construction process. 
This previous work aims at discovering concepts from a pre-specified set of event definitions within a known domain and studying the large beneficial impacts of such event-specific concepts in event retrieval, zero-example detection, and summarization. In contrast, this paper focuses on discovering semantic concepts from an external event knowledge base, WikiHow, and using its rich ontological hierarchy to organize the large number of concepts learned from the Web. Such ontological structure plays an important role
in handling novel events that have not been seen in the learning stage, as confirmed by the significant performance gains reported in the experiments.

There are some existing concept libraries built for different purposes.
Object bank \cite{object_bank} consists of $200$ objects taken from the intersection set of most frequent $1,000$ objects between image datasets LabelMe \cite{labelme} and ImageNet \cite{imagenet}. Each object detector is trained with $100\sim 200$ images and their object bounding boxes.
Classemes \cite{classemes} is a concept library comprised of $2,659$ concepts defined from Large Scale Concept Ontology for Multimedia (LSCOM) \cite{LSCOM}.
Each concept detector is trained with $150$ images from \emph{bing.com} search engine using the LP-$\beta$ multiple kernel learning algorithm.
Action bank \cite{action_bank} is built for high-level representation of human activities in video, which contains $205$ template actions.
Although these libraries achieve good performances on different tasks, they are not designed for video events.
In our work, we build a concept library specifically designed for video events, and the concept ontology and models are automatically discovered from the Web.

\newpage
\section{Building Concept Bank Ontology}
\label{sec_build}
In this section, we will introduce the procedure of building Concept Bank ontology, including event collecting and concept discovery for each event.

\subsection{Collecting Events from WikiHow}
\label{sec_wikihow}
\begin{figure} [!ht]
\centering
\includegraphics[width=1\columnwidth]{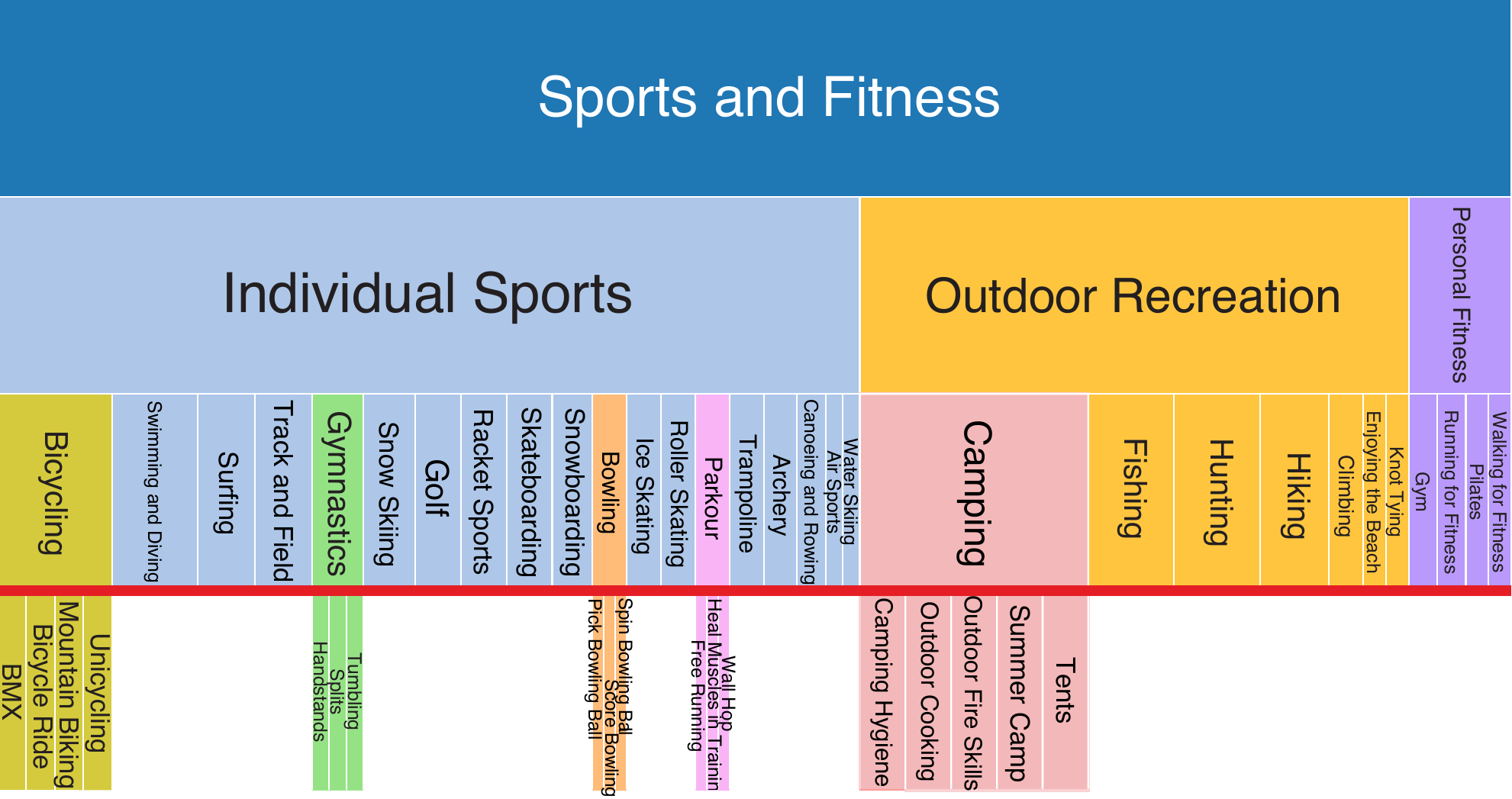}
\caption{ A portion of hierarchical structure for event category ``Sports and Fitness'' in WikiHow. }
\label{fig_wikihow}
\end{figure}

In order to get a quasi-comprehensive list of events, we choose among the events from WikiHow \cite{wikihow}.
WikiHow is a wiki, similar to Wikipedia, in which internet users can read or edit the articles.
Currently, WikiHow contains $163,957$ articles that are organized into $20$ categories in a hierarchical structure.
The $20$ categories cover the major aspects of human daily life including ``Food and Entertaining'', ``Pets and Animals", ``Sports and Fitness" and so on.
Figure \ref{fig_wikihow} illustrates a portion of the hierarchical structure of event category ``Sports and Fitness'' in WikiHow.
Each category contains a number of subcategories (\emph{e.g.}, ``Individual Sports" and ``Outdoor Recreation") that are instantiations of the higher level event catalog.  Subcategories are further divided into various fine-grained events in hierarchy (\emph{e.g.}, ``Bicycling", ``Mountain Biking" and ``Fishing''), each has several articles (\emph{e.g.}, ``Jump a Mountain Bike" under ``Mountain Biking"), which are not shown in the figure, describing the detailed procedure of each specific event.

We find that some events in WikiHow are closely related and are sub-events within an event (\emph{e.g.}, ``Mountain Biking'' and ``Unicycling" under ``Bicycling").
Our intuition is that organizing events into such fine granularity will result in a huge event collection with heavy semantic redundancy, which is impractical and unnecessary for our task. Therefore, we only focus on the first three layers in WikiHow (above the bold solid red line in Figure \ref{fig_wikihow}) and treat all events in the third layer as potential events in our library, resulting in $1,257$ event candidates in total.

Furthermore, we notice that some events are not visually detectable.
For example, events ``getting good grades in college'' or ``being successful'' do not convey consistent visual patterns that can be modeled by computer vision techniques.
Therefore, a filtering is performed to remove such undetectable events.
To this end, we ask two human labelers to judge whether each of the $1,257$ events corresponds to a visually detectable event.
After filtering, we end up with $631$ events which are included as the events in our library.


\subsection{Concept Discovery for Each Event}
\label{sec_flickr}
After collecting events from WikiHow, we need to associate a number of concepts to each event.
One straightforward way is to use advanced Natural Language Processing technique to extract concepts from article contents on WikiHow.
However, besides the complexity of the procedure itself, the resultant concepts do not correspond to any visual patterns, which also brings challenges for training data acquisition. On the other hand, the images on Flickr are associated with tags that describe the semantic of visual content in different events. Therefore, we leverage the tags of Flickr images to discover potential concepts for events.

In this work, we adopt our previous work on event-driven concept discovery from Internet images~\cite{FlickrConcept} to discover concepts for WikiHow events. For each of the $631$ events, we use the article names under this event as textual queries to crawl images and their associated tags from \emph{Flickr.com} image search engine. We only keep color images with resolution higher than $200 \times 200$ pixels.
The crawling ends up with $727,910$ images in total and the average number of images is $1,154$ per event.
Since some tags are not semantically meaningful words (such as a camera brand name) or do not have consistent visual patterns (such as ``biology'' or ``economy''), directly accepting these tags as concepts may introduce lots of noises and degrade the performance.
Therefore, we adopt the following two-phase process to discover concepts that have both semantic meanings and visually consistent patterns from tags.

\textbf{Step I : Noisy Tag Cleansing.}
Notice that a tag could be either a word or a phrase.
So firstly, we convert a tag into lower case and divide the tag into tokens by a tokenizer, then remove non-word tokens and tokens belonging to stop words (like ``the'', ``and'', or ``with'') \cite{nltk}.
We then define a meaningless-word-list including camera names (like iPhone or Nikon) and people names (like Peter or Julia).
For a tag, if one of its tokens appears in our meaningless-word-list or does not match any synonyms in WordNet, the tag is regarded as meaningless.
During this step, we also bundle together tokens within a WordNet synonym with lemmatization \cite{nltk}.
For example, ``run'', ``running'', ``runs'' and ``ran'' will be regarded as a same token ``run".
By doing this, only the tags with semantic meaning are retained.
Finally, we rank tags within each event in descending order by their frequencies and select top $100$ tags as candidate concepts.

\textbf{Step II : Concept Visualness Verification.}
If a concept is not visually related, the concept detector will be hard to generalize due to high intra-class variance \cite{auto_attributes}.
Therefore, we use the cross-validation performance as the measure of visualness.
Specifically, for a candidate concept, we collect images with the concept as positive samples and randomly select a same number of images which do not contain the concept as negative samples.
Then we build a concept detector (see Section \ref{sec_classifier} for details) and calculate the 2-fold cross-validation performance. In this procedure, we adopt Average Precision (AP) as the performance evaluation metric.
Finally, only candidate concepts with cross-validation APs higher than $0.8$ are selected as concepts for the event.

In order to get statistically reliable concept detectors, we further do a post processing that removes those concepts contain less than $100$ training images.
By doing this, we end up with $4,876$ concepts.
Note that the same concept may appear in different events (and are trained with different training images, see Section~\ref{sec_classifier}). Therefore, we call our discovered concepts as ``event-specific concepts''.

\subsection{Concept Bank Ontology}
\label{sec_structure}
\begin{figure} [!ht]
\centering
\includegraphics[width=1\columnwidth]{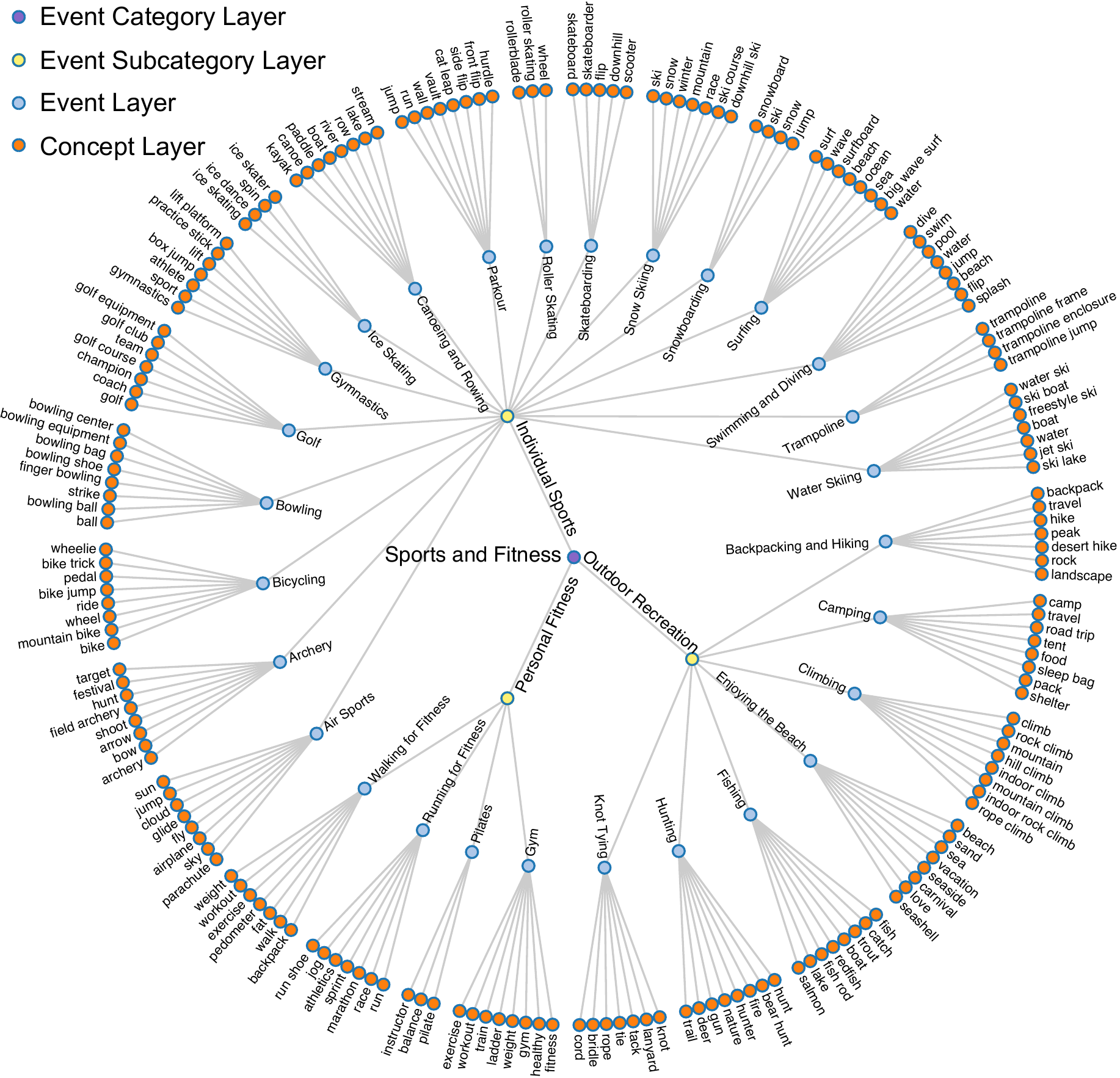}
\caption{A branch of the hierarchy structure in Concept Bank ontology corresponding to event category ``Sports and Fitness''.}
\label{fig_structure}
\end{figure}

Based on the collected events and concepts, we construct a five-layer tree structured concept ontology that helps event detection (see Section~\ref{sec_representation}).
In more details, the root layer is a single node covering all events and concepts in the Concept Bank.
The second layer consists of $19$ event categories ranging from ``Arts and Entertainment'' to ``Travel'', each corresponds to a category in WikiHow.
In the third layer, we have in total $130$ nodes in which each node corresponds to an event subcategory.
In the fourth layer, we have $631$ events.
Furthermore, each of the $631$ event nodes has several child nodes corresponding to its event-specific concepts.
Therefore, the bottom layer of the CB tree ontology contains $4,876$ leaf nodes (concepts).
Figure \ref{fig_structure} illustrates a branch of hierarchy structure in CB ontology corresponding to event category ``Sports and Fitness''.
Due to the space limitation, we remove the root node and only show at most $8$ concepts for each event.
In the next section, we will associate a concept detector with each of the leaf nodes.

\newpage

\newpage
\section{Learning Concept Detector}
\label{sec_classifier}
In this section, we will discuss how to build concept detector for each discovered concept.

\subsection{Feature Extraction}
\label{sec_feature}
In order to describe visual content information comprehensively and complementarily,
we select $5$ low-level feature descriptors that are proven to be effective in representing visual content. They are respectively
\textbf{SIFT} \cite{sift}, \textbf{GIST} \cite{gist}, \textbf{Gabor} \cite{gabor}, \textbf{LBP} \cite{lbp} and \textbf{Transformed Color Distribution} \cite{color}.
All descriptors are densely extracted on grids of $20 \times 20$ pixels with $50\%$ overlap from images.
For each type of the extracted descriptors, we train a codebook with $1,000$ codewords, and partition each image
into $1 \times 1$, $2 \times 2$ and $3 \times 1$ blocks for spatial pyramid matching \cite{spm}. Finally, we adopt
soft quantization \cite{soft_quantization} to represent each image as a $8,000$ dimensional histogram for each feature.
We also adopt these features in generating concept scores for videos.

\subsection{Training Image Selection}
\label{sec_image_selection}
Since images and their tags are crawled from the web, images associated with a concept are noisy and contain some outliers.
Therefore, we need to remove the noisy images before training concept detectors.
Our intuition is that the majority of images in a concept form a common ``visual theme'' in the image search result, while distracting images tend to depart from such visual commonality. This motivates us to develop a collective approach to measure the relative confidence of each image in the search result.
Specifically, for each concept $c$, we collect a set of images $F=\{1,\ldots,N\}$ that contain the concept as their tag, where $N$ is the total number of images.
For image $i$, we extract $M$ features ($M=5$ in this work) and denote its feature vector as $\mathbf{f}_{m}^{i}\in\mathbb{R}^{d_m}$ ($m=1,\ldots,M$) with $d_{m}$ denoting the feature dimensionality of the $m$-th feature. We adopt Kernel Density Estimation (KDE) method \cite{kde} to estimate the probability of image $i$ belonging to concept $c$ defined as:
\begin{equation}
p(c | i) = \frac{1}{M\times N} \sum_{j=1}^{N} \sum_{m = 1}^M \mathcal{G}_m (\mathbf{f}_m^i - \mathbf{f}_{m}^{j}),
\end{equation}
where $\mathcal{G}_m(\cdot)$ is the Gaussian kernel function of the $m$-th feature defined as:
\begin{equation}
\mathcal{G}_m (\mathbf{f}_{m}^i - \mathbf{f}_{m}^j) = \exp (-\frac{\| \mathbf{f}_{m}^i - \mathbf{f}_{m}^j \|^2}{\sigma_m^2}),
\end{equation}
in which $\|\cdot\|$ denotes the $l_2$ norm, $\sigma_m$ is the kernel radius of the $m$-th feature setting as the mean value of all pairwise distance among the images.

\begin{figure} [!ht]
\centering
\includegraphics[width=1\columnwidth]{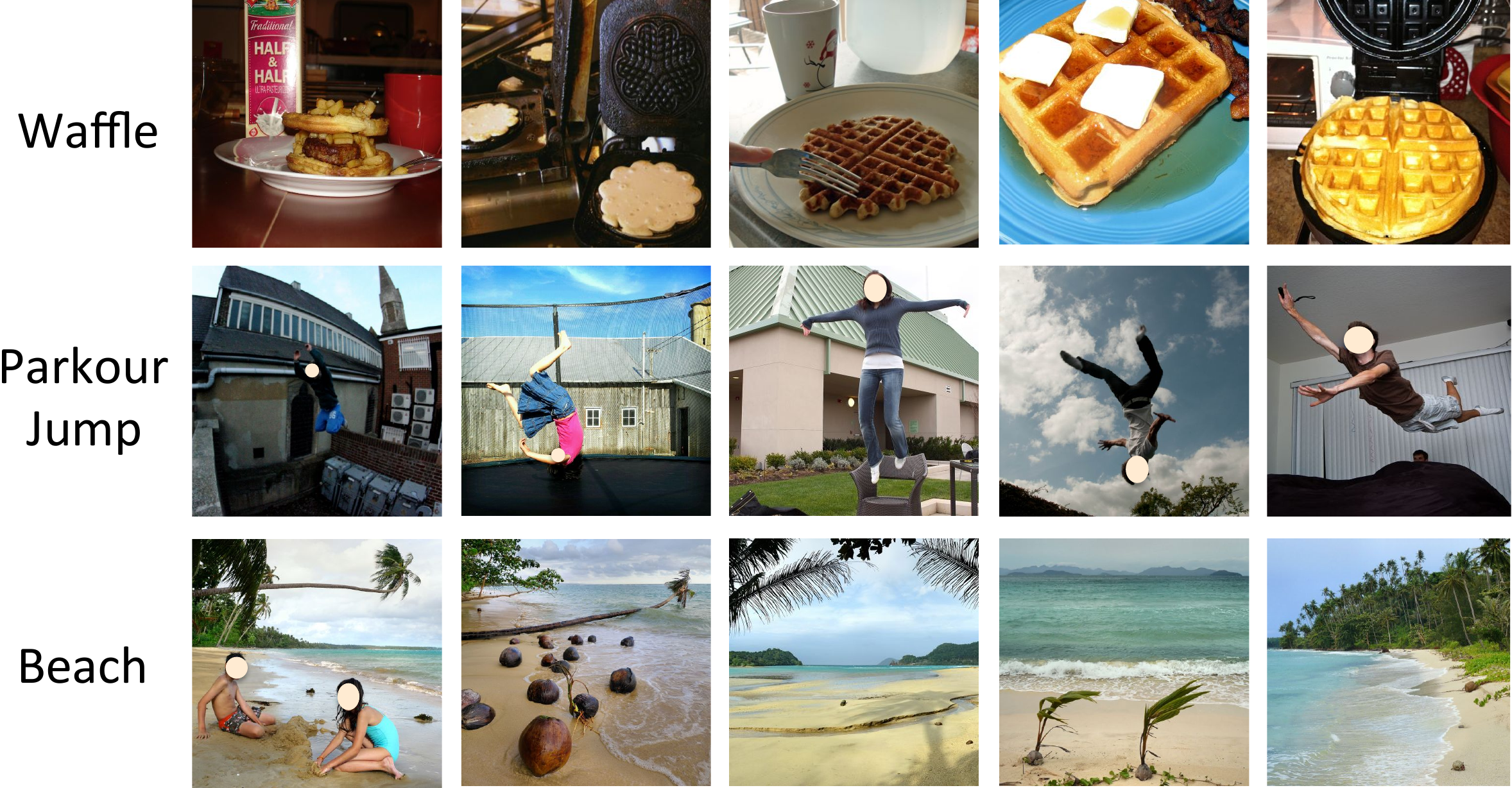}
\caption{Top $5$ selected Web images for $3$ exemplary concepts using our training image selection method.}
\label{fig_noise}
\end{figure}

We use $p(c|i)$ as the confidence score of image $i$ belonging to concept $c$ and select $s$ images with highest confidence scores as the positive training samples for each concept.
In this work, we set $s=100$ and choose $t=1,000$ images from other concepts as negative training samples.
Figure~\ref{fig_noise} shows $5$ top selected images for exemplary concepts, from which we can see that the selected images are highly reliable while at the same time achieve reasonable diversity.

Since a concept may appear under different events, our method will select different training images for each event-specific concept detector. Therefore, each concept detector will convey context information for its belonging event.

\subsection{Concept Detector Training}
After selecting $s$ positive and $t$ negative training images for each concept, we use Multiple Kernel Linear SVM (MK-LSVM) classifier with L1-Loss to train our concept detector. The dual norm of its objective function is:
\begin{eqnarray}\nonumber
\min_{\boldsymbol{\alpha}, \boldsymbol{\beta}} && \frac{1}{2} \boldsymbol{\alpha}^T \big( \sum_{m=1}^M \beta_m \mathcal{K}_m + \gamma I \big) \boldsymbol{\alpha} - \mathds{1}^T \boldsymbol{\alpha} \\ \nonumber
\mathrm{s.t.} && \boldsymbol{\alpha}^T \boldsymbol{y} = 0,\ 0 \leq \alpha_i \leq C,~~ \ \forall i, \\
&& \sum_{m=1}^M \beta_m = 1,\ \beta_m \geq 0,~~ \ \forall m,
\end{eqnarray}
where $\mathbf{y}=[y_1,\ldots,y_{s+t}]^{\top}\in\{0,1\}^{s+t}$ is the label vector with $y_i=1$ if the $i$-th image is positive and $y_i=0$ otherwise,
$\boldsymbol{\alpha} = [\alpha_1,\ldots,\alpha_{s+t}]^{\top}\in\mathbb{R}^{s+t}$ is the dual variable vector, $\boldsymbol{\beta}=[\beta_1,\ldots,\beta_M]^{\top}\in\mathbb{R}^{M}$ is the kernel weight vector, $\mathcal{K}_m\in\mathbb{R}^{(s+t)\times(s+t)}$ is the Gram matrix of the $m$-th feature, $I$ is an identity matrix , $\mathds{1}$ is the all-one vector, $\gamma$ and $C$ are model parameters.
We use LIBLINEAR \cite{liblinear} and choose $\gamma = 0.01$ and $C=1$ in training concept detectors.

%

\newpage
\section{Representing Videos with Concept Bank}
\label{sec_representation}
Given a video clip, we can directly apply all $4,876$ concept detectors in Concept Bank to generate the concept representation.
However, as aforementioned, it is desirable to choose the most relevant concepts related to the query event and generate a compact concept representation.

In our work, we turn to use the semantic similarity matching between event and concept names to find relevant concepts for an event query.
In order to determine semantic similarity, we adopt ConceptNet~\cite{conceptnet}, a semantic network developed by MIT Media Lab containing common sense knowledge for computers to capture human world. We use Divisi, a Python API for ConceptNet, to calculate semantic similarity between two words.
The similarity between two phrases are the maximum similarity between all pairs of words in the two phrases.

Denote by $sim(a, b)$ the similarity between two words (phrases) $a$ and $b$ calculated from ConceptNet.
For a given event $e$ and a concept $c_0$, we take advantage of the hierarchy in the Concept Bank ontology to calculate their semantic similarity.
Specifically, denote $c_1$, $c_2$ and $c_3$ as the three ancestors of concept $c_0$ in Concept Bank ontology residing at event layer, event subcategory layer and event category layer respectively.
Then the semantic similarity $S(e,c_0)$ between $e$ and $c_0$ can be calculated as follows:
\begin{equation}\label{Eq:HieSim}
S(e,c_0) = \prod_{i=0}^3 sim(e, c_i).
\end{equation}
For instance, for event ``grooming a dog'', we calculate its similarity with concept ``brush'' (similarity value $0.7$) and its three ancestors: , ``dog grooming'' (similarity value $1$), ``dogs'' (similarity value $1$), and ``pets and animals'' (similarity value $0.8$). Then the similarity between event ``grooming a dog'' and concept ``brush'' is $0.8 \times 1 \times 1 \times 0.7 = 0.56$.

Eq. (\ref{Eq:HieSim}) can be justified as follows. Since the same concept may appear under different events, involving high-level ancestors in the similarity calculation
will measure the relevance of each concept based on the coherent similarity of its ancestors with respect to the event query, leading to more precise relevant concept selection. In this way, when answering event query ``grooming a dog'', we can effectively select concept ``brush'' from event ``dog grooming'' rather than from event ``personal makeup'' under ``Personal Care and Style" even if it also contains concept ``brush''.


With semantic similarity matching, we choose the top $n$ most relevant concepts for the query event and use their concept detectors to generate concept based video representation.
Specifically, we evenly sample $m$ frames from a video clip.
Then we apply the selected concept detectors on each frame and adopt their probabilistic outputs as concept scores.
The final representation of the video is the average of concept score vectors across the sampled frames.
We choose $n=100$ and $m=20$ in the experiments.
%

\newpage
\section{Experimental Evaluations}
\label{sec_experiment}
In this section, we evaluate the effectiveness of the Concept Bank in video event detection.
We first introduce the datasets and comparison methods and then present experiment analysis on two event detection tasks.
Specifically, we would like to evaluate whether the large number of concepts discovered using WikiHow events can be used to effectively detect events specified in some benchmark datasets.
Additionally, we aim at comparing different ways of training the concept detectors and mapping events to concepts.

\subsection{Dataset and Comparison Methods}
\textbf{Dataset}.
We evaluate our proposed CB representation on two video event detection datasets:
(1) \textbf{TRECVID 2013 Multimedia Event Detection (MED) dataset}.
This is a dataset of $32,744$ videos over $20$ event categories (i.e., the partition used in the Pre-Specified EK$100$ task~\cite{MED}).
These $20$ event categories are (from E1 to E20 respectively):
``\emph{birthday party}", ``\emph{changing a vehicle tire}", ``\emph{flash mob gathering}", ``\emph{getting a vehicle unstuck}",
``\emph{grooming an animal}", ``\emph{making a sandwich}", ``\emph{parade}", ``\emph{parkour}", ``\emph{repairing an appliance}",
``\emph{working on a sewing project}", ``\emph{attempting a bike trick}", ``\emph{cleaning an appliance}", ``\emph{dog show}",
``\emph{giving directions to a location}", ``\emph{marriage proposal}", ``\emph{renovating a home}", ``\emph{rock climbing}",
``\emph{town hall meeting}", ``\emph{winning a race without a vehicle}", ``\emph{working on a metal crafts project}".
We follow the standard partition of this dataset, which includes $7,787$ videos in the training set and $24,957$ videos in the test set. Achieving good performance on this dataset is quite challenging because the majority of videos in this dataset are background videos
(only $2,000$ videos in training set and $1,489$ videos in test set belong to one of the $20$ events).
(2) \textbf{Columbia Consumer Video (CCV) dataset}.
This is a dataset of $9,317$ YouTube videos over $20$ event categories, where $4,659$ videos are used for training and $4,658$ videos for testing \cite{ccv}.
These $20$ categories are (from E1 to E20 respectively):
``\emph{basketball}", ``\emph{baseball}", ``\emph{soccer}", ``\emph{ice skating}", ``\emph{skiing}", ``\emph{swimming}",
``\emph{biking}", ``\emph{cat}", ``\emph{dog}", ``\emph{bird}", ``\emph{graduation}", ``\emph{birthday}", ``\emph{wedding reception}",
``\emph{wedding ceremony}", ``\emph{wedding dance}", ``\emph{music performance}", ``\emph{non-music performance}",
``\emph{parade}", ``\emph{beach}", ``\emph{playground}".

\textbf{Comparison Methods}. We compare the concept based video representations generated from the following methods.
(1) \textbf{Classemes} \cite{classemes}, a $2,659$ dimensional concept representation built on the LSCOM concept ontology~\cite{LSCOM}. We generate Classemes feature from each video frame and then average all frame features across a video as the video-level feature.
(2) Selected Concepts from Best single Feature (\textbf{SCBF}).
The concept detector is trained based on the best performing single feature, in which the training images are selected with the method described in Section~\ref{sec_image_selection}.
When a query event comes, it uses the semantic matching method in Section~\ref{sec_representation} to select relevant concepts for video representation.
(3) Selected Concepts from Random Images (\textbf{SCRI}). The concept detector is trained by MK-LSVM, but the training images are randomly selected. The relevant concepts are selected using semantic matching.
(4) Random Concepts from Selected Images (\textbf{RCSI}). The concept detector is trained by MK-LSVM with selected training images. However, concepts for the query event are randomly selected, instead of using semantic matching.
(5) Our Selected Concepts from Selected Images (\textbf{SCSI}). The concept detector is trained by MK-LSVM with selected training images. The relevant concepts are selected based on semantic matching.

\textbf{Evaluation Tasks}. For each method, we evaluate the performance of different representations under two video event detection tasks: Event Detection in Concept Space and Zero-Shot Event Retrieval. In each task, the Average Precision (AP), which approximates the area under precision/recall curve, is used as evaluation metric on each event. We calculate mean Average Precision (mAP) over all event categories as the overall evaluation metric on the dataset.

\subsection{Event Detection in Concept Space}
\label{sec_ed}
In this task, all representations are regarded as high-level descriptors in concept space for video event detection.
On each dataset, we choose $100$ concepts for each of $20$ events from the Concept Bank and concatenate the concept scores into a $2,000$-dimensional feature vector.
A one-vs-all SVM classifier with $\chi^2$ kernel is trained as the event detector.
In the test stage, we use SVM probabilistic output as the event confidence score on each test video.
Best parameters for SVM are chosen via $2$-fold cross-validation on training set.

\begin{figure}[!ht]
\centering
\subfigure
{\label{fig_ed:a}
\includegraphics[width=1\columnwidth]{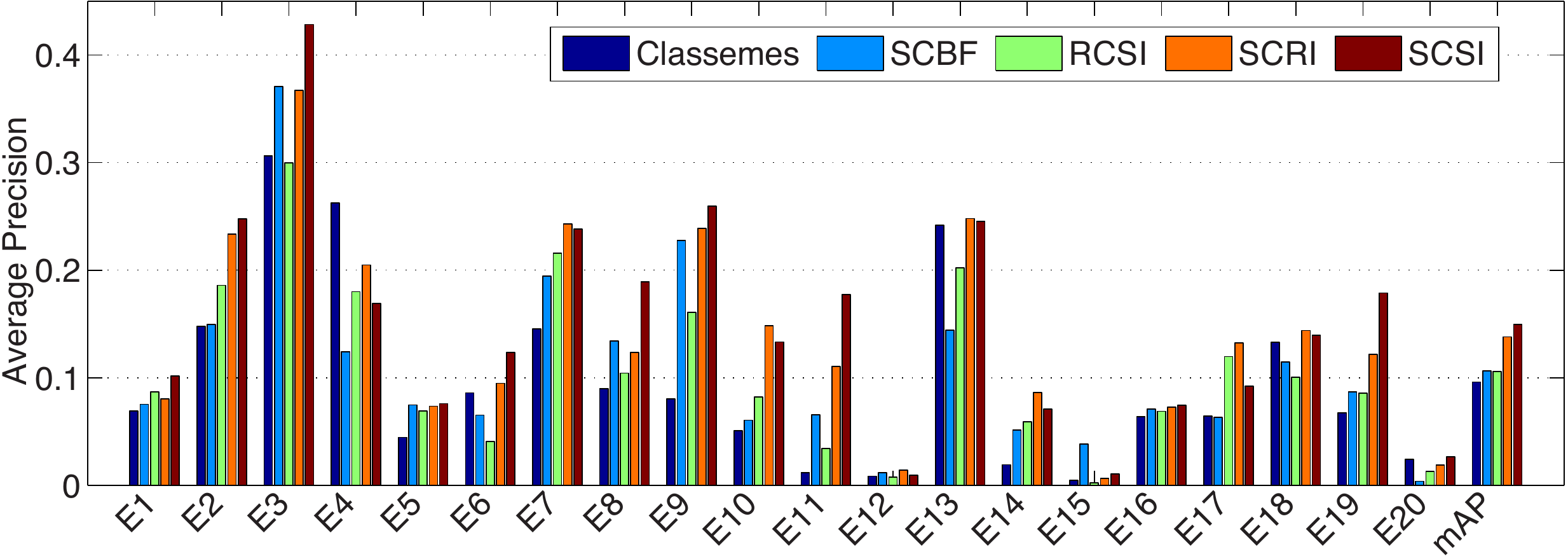}
}
\subfigure
{ \label{fig_ed:b}
\includegraphics[width=1\columnwidth]{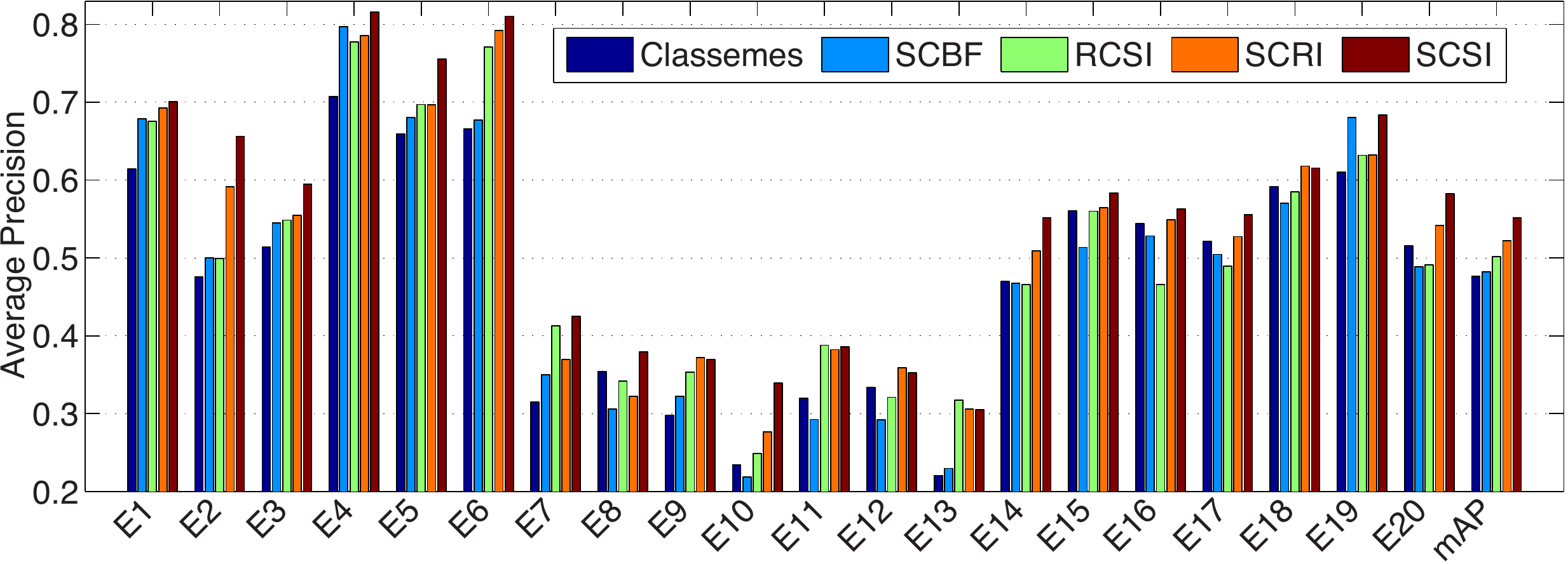}
}
\caption{ Performance comparison on event detection in concept space task (left: MED; right: CCV). }
\label{fig_ed}
\end{figure}

The comparison results between the five methods on MED and CCV are given in Figure \ref{fig_ed}, from which we have following observations:
(1) The proposed SCSI method achieves the best performance by a significant margin compared with other methods, which demonstrates the effectiveness of SCSI in generating
reliable concept based video representations.
(2) The multiple feature based methods including SCSI and SCRI clearly beat the single feature based method SCBF. This shows the advantages of utilizing multiple features in concept detector training.
(3) The SCSI significantly outperforms RCSI, which verifies that our semantic similarity matching method is able to select relevant concepts for event representation.
(4) The SCSI has higher performance than SCRI. This is due to the fact that the former leverages more reliable training images in concept detector training than the random images utilized in the latter.

\begin{figure}[!ht]
\centering
\subfigure
{\label{fig_ed_noc:a}
\includegraphics[width=0.466\columnwidth]{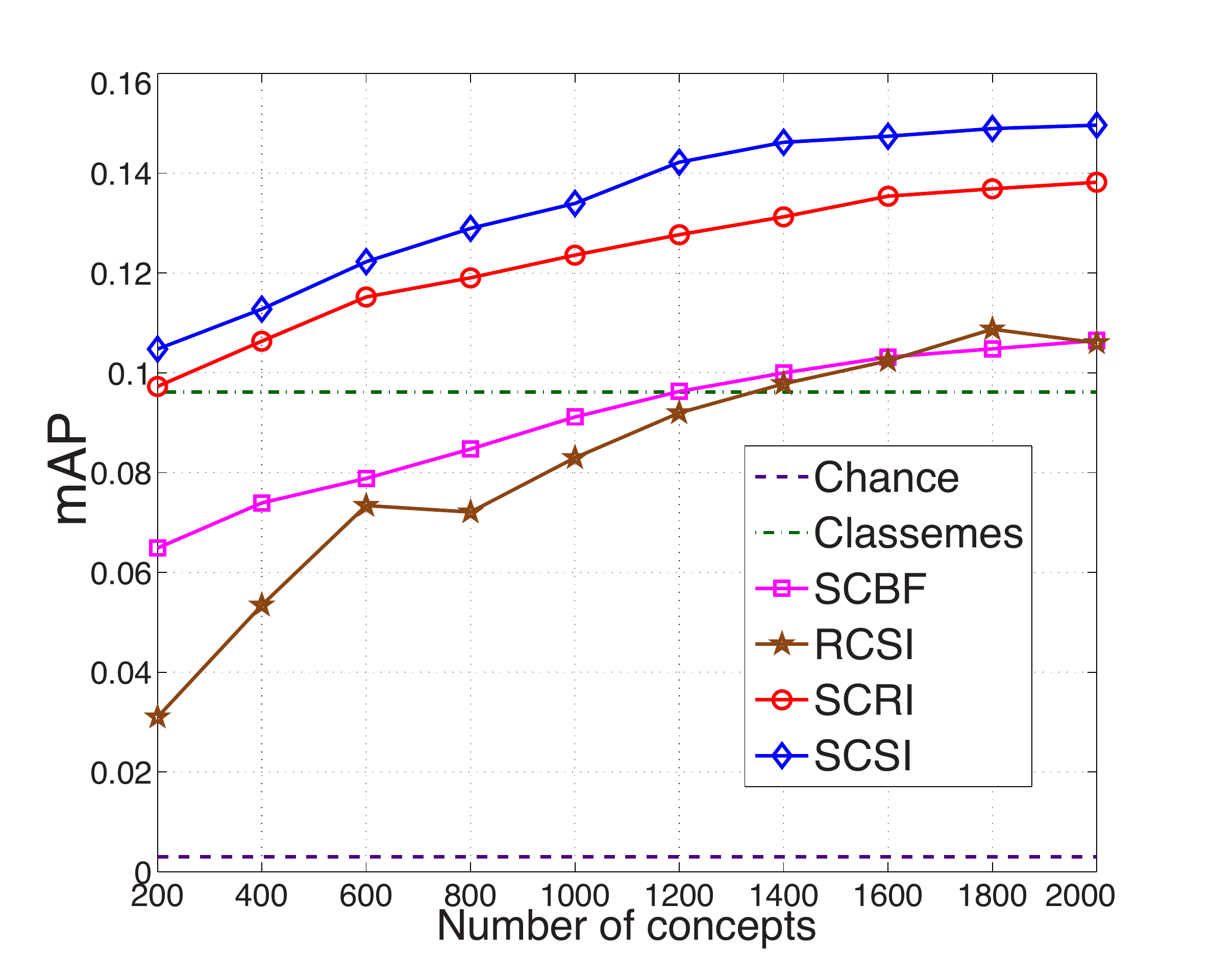}
}
\subfigure
{ \label{fig_ed_noc:b}
\includegraphics[width=0.482\columnwidth]{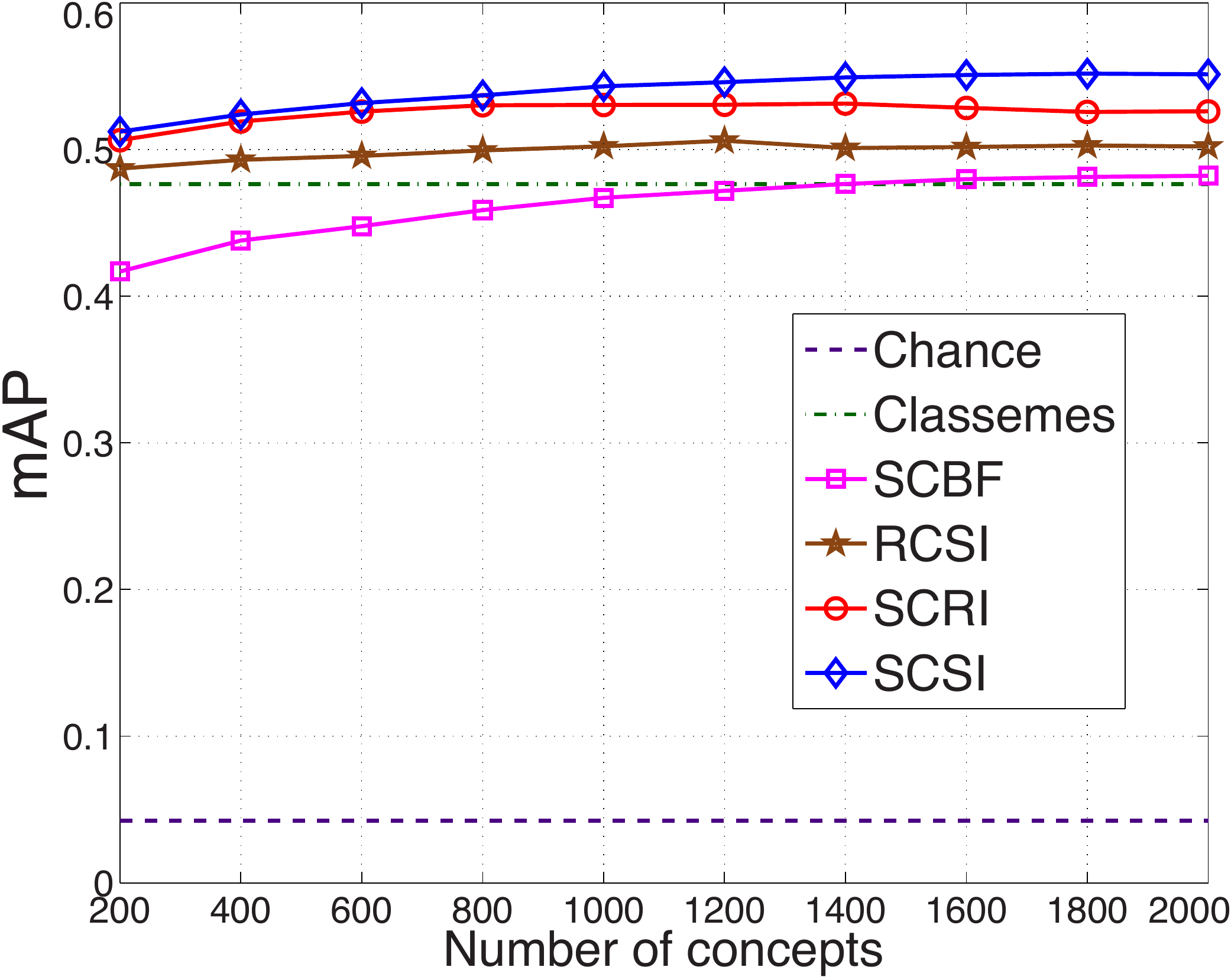}
}
\caption{ Event detection performance with different number of concepts (left: MED; right: CCV). }
\label{fig_ed_noc}
\end{figure}

Figure \ref{fig_ed_noc} illustrates the variation of mAP with different number of concepts.
For SCSI, SCRI, SCBF and RCSI, we choose the number of concepts from $10$ to $100$ with step of $10$ for each event, and get concept numbers rang from $200$ to $2,000$.
For Classemes, we fix the concept dimension to be $2,659$ (lower dimension of classemes leads to worse performance).
As seen, the proposed SCSI method consistently outperforms all other methods when different numbers of concepts are adopted in each of the two datasets.
This clearly demonstrates that our method is able to generate accurate concept scores on the videos to facilitate concept based event detection.
Furthermore, the event detection performance keeps improving as the number of concepts increases.
This shows that event modeling over concept space can benefit from the increasing number of relevant concepts.
We also notice that, on both MED and CCV datasets, the proposed SCSI outperforms $2,659$ dimension Classemes even with only $200$ dimensions ($10$ concepts for each of the $20$ events), which again clearly demonstrates the advantages of our CB representation.

\subsection{Zero-Shot Event Retrieval}
\label{sec_er}
In this task, we do not use any training videos, but directly use the concept scores on the test video to perform event retrieval.
We call this zero-shot event retrieval since the procedure is purely semantic based without using any training examples.
Specifically, we rank all test videos based on the detection scores of each selected relevant concept and then fuse the rank lists of all relevant concepts to generate the final rank list.
In this work, we adopt the normalized rank fusion method \cite{rank_fusion} to combine multiple ranking lists.
Specifically, for a test set with $n$ videos, the fusion score $R(i)$ for $i$-th video is calculated by $R(i) = \frac{1}{d} \sum_{j=1}^d ( 1 - \frac{r_j}{n} )$,
where $r_j \in \{1, 2, \dots, n\}$ is the rank position of $i$-th video in the rank list generated by $j$-th concept and $d$ is the concept number.
In this way, we get rid of the influence caused by numeric scale differences of raw concept scores generated from different concept detectors.
Finally, we rank test videos based on the fusion scores.

In zero-shot retrieval, since the only available information is the query event name, we do not utilize concept names from other event categories.
Therefore, we only choose $100$ relevant concepts selected for the specific query event based on semantic similarity matching described in Section \ref{sec_representation}.
To get a fair comparison, we also select $100$ concepts from Classemes using the same method.

\begin{figure}[!ht]
\centering
\subfigure
{\label{fig_sr:a}
\includegraphics[width=1\columnwidth]{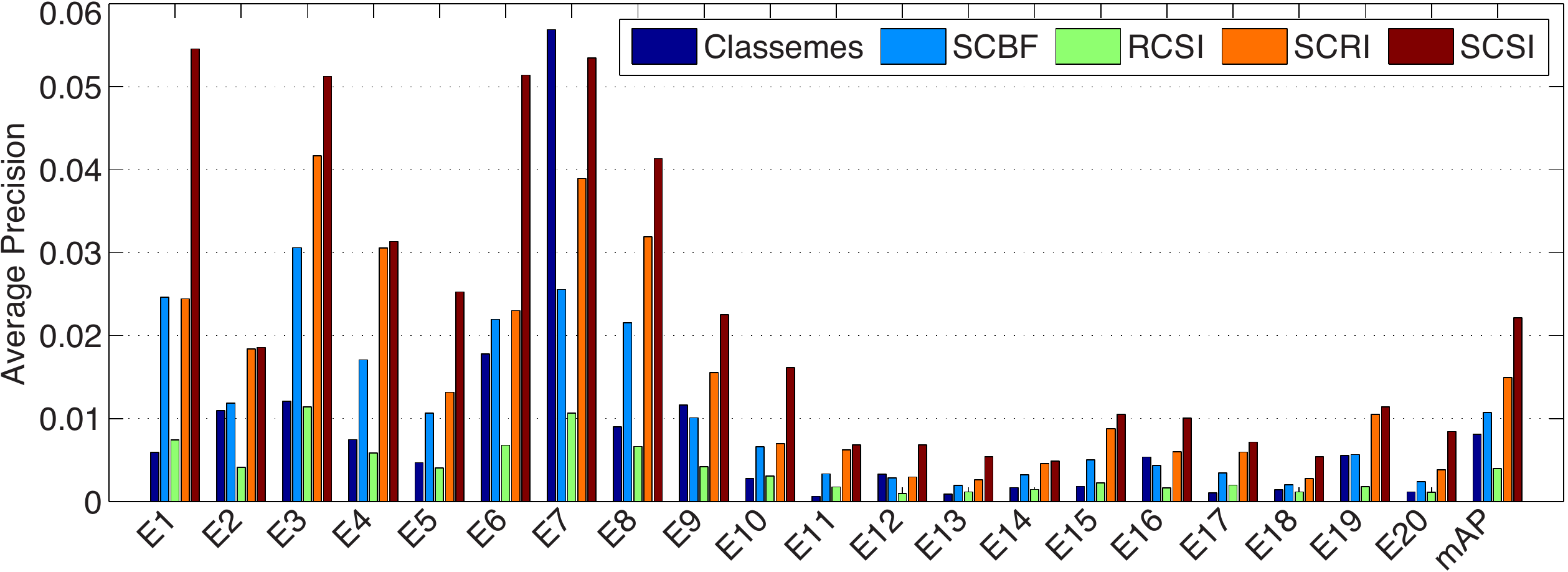}
}
\subfigure
{ \label{fig_sr:b}
\includegraphics[width=1\columnwidth]{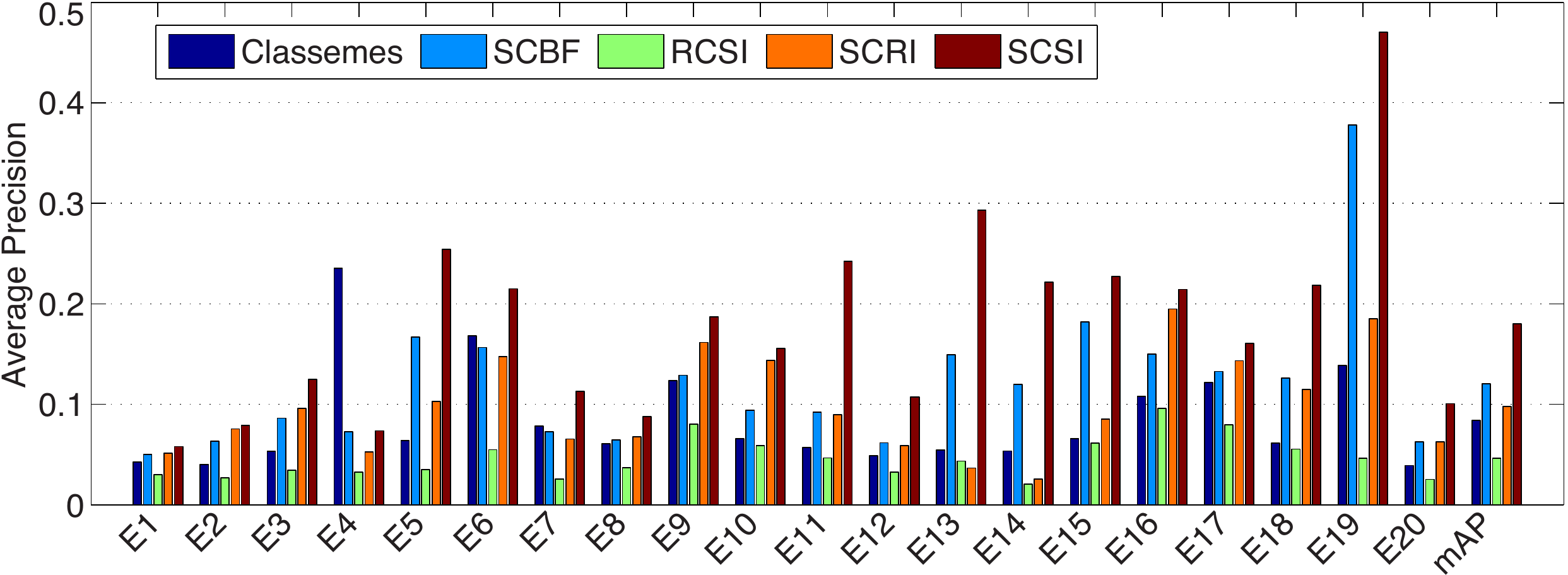}
}
\caption{ Performance comparison on zero-shot event retrieval task (left: MED; right: CCV). }
\label{fig_sr}
\end{figure}

\begin{figure}[!ht]
\centering
\subfigure
{\label{fig_sr_noc:a}
\includegraphics[width=0.474\columnwidth]{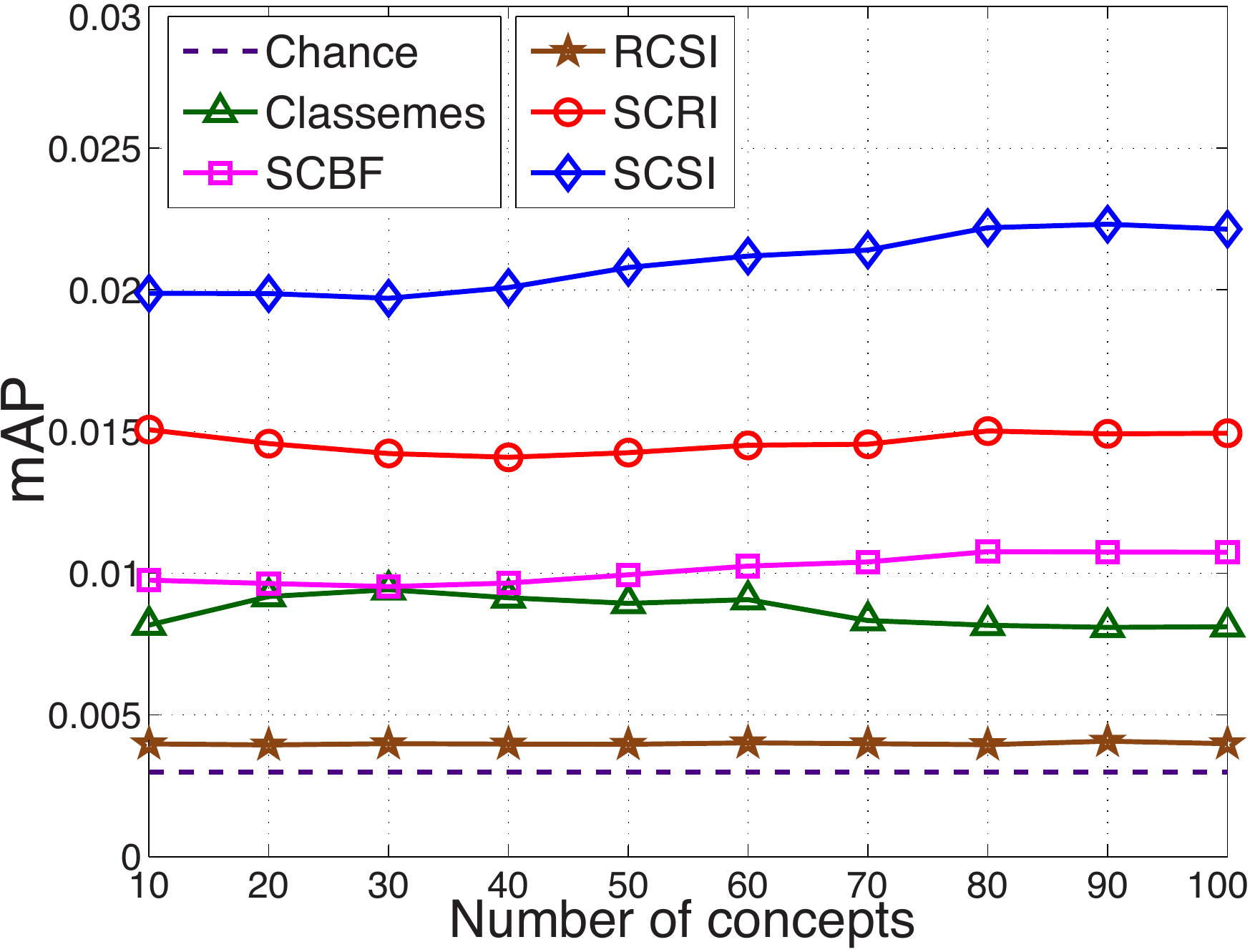}
}
\subfigure
{ \label{fig_sr_noc:b}
\includegraphics[width=0.474\columnwidth]{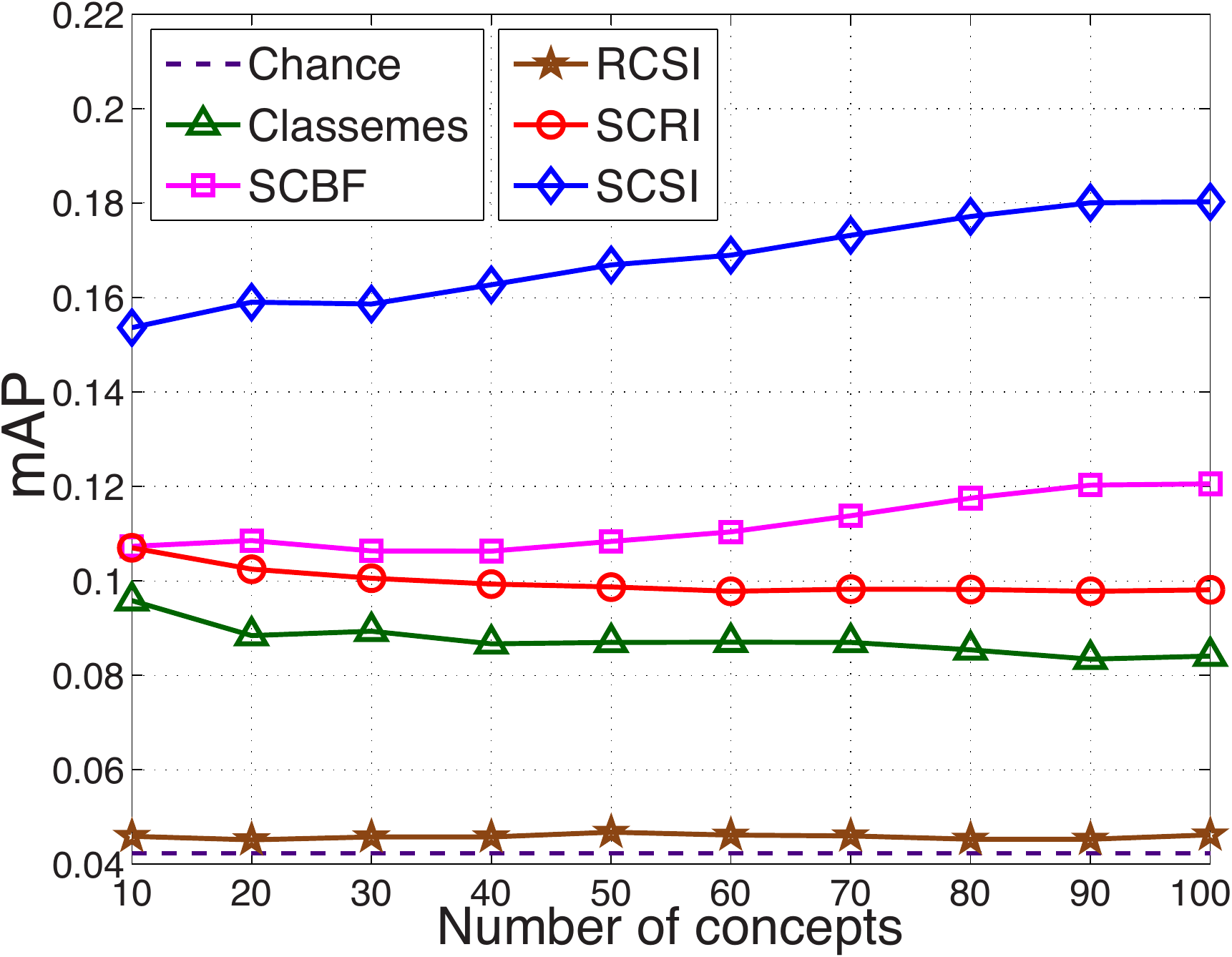}
}
\caption{ Zero-shot event retrieval performance with different number of concepts (left: MED; right: CCV). }
\label{fig_sr_noc}
\end{figure}

Figure \ref{fig_sr} shows the comparison results between the five methods for zero-shot retrieval task and Figure \ref{fig_sr_noc} illustrates the variation of mAP with different number of concepts.
There are two points we want to mention:
(1) The performance of zero-shot event retrieval is worse than that of supervised event modeling over concept space (Section \ref{sec_ed}). This is because the latter uses training samples to obtain a more sophisticated event model while the former is merely based on semantic score fusion.
Notably, since the concepts are all randomly chosen and irrelevant to the query event, the performance of RCSI degrades a lot and is close to chance in this task.
(2) Our SCSI method keeps improving as the number of concepts, whereas the performance of Classemes decreases.
This indicates that Classemes does not have enough concepts related to the event, so the increasing number of irrelevant concepts will not help the detection. However, our Concept Bank has a broader coverage thus it can benefit from the increasing number of concepts.

\newpage
\subsection{Semantic Recounting in Videos}
\begin{figure}[!ht]
\centering
\includegraphics[width=1\columnwidth]{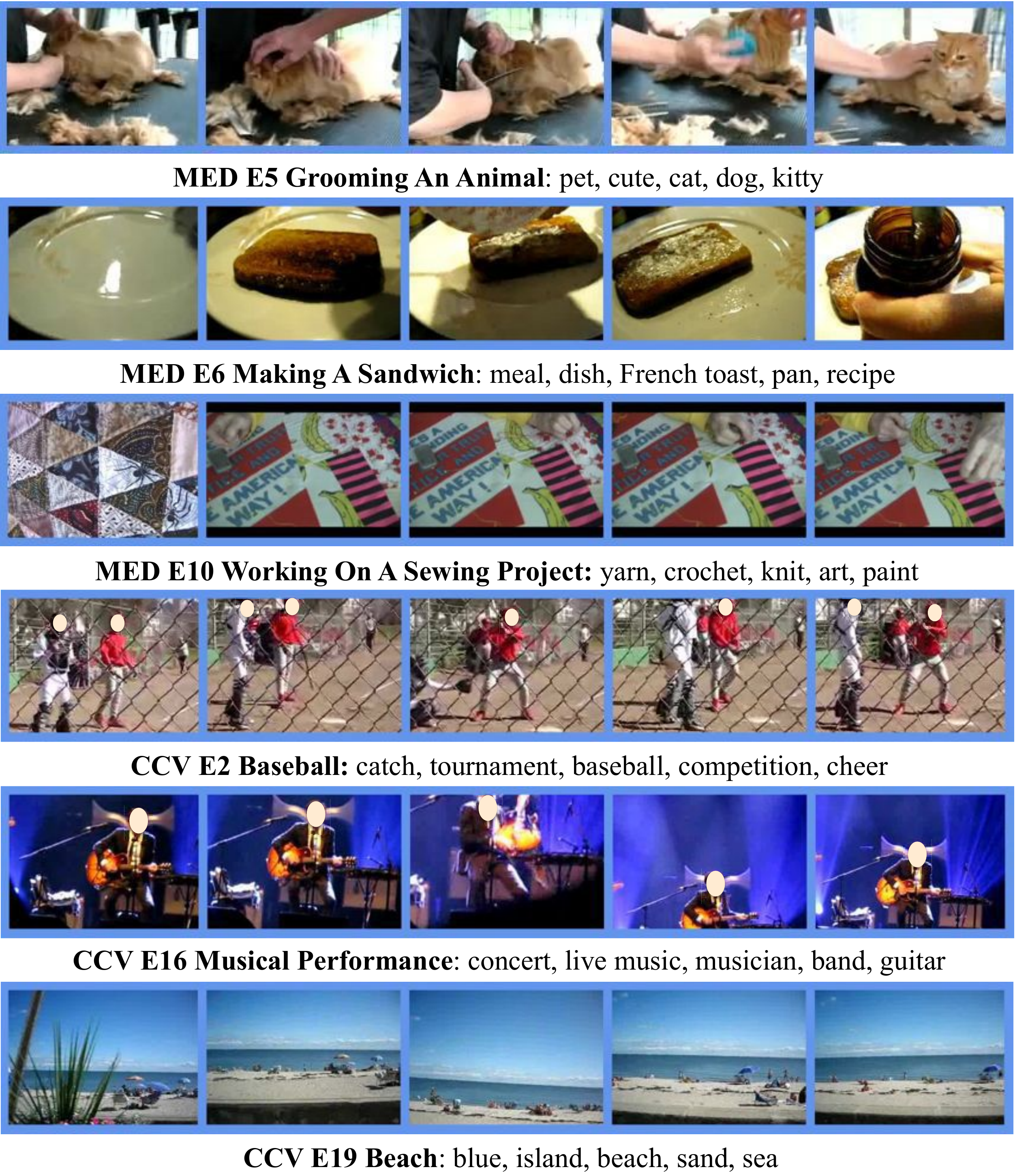}
\caption{ Event Video Recounting Results: each of the $6$ rows shows evenly subsampled frames of an example video and the top $5$ relevant concepts detected in the video. }
\label{fig_recounting}
\end{figure}

For each video of a target event, we rank all the concepts discovered for the event based on their confidence scores and treat the top ranked concepts as
the semantic description of the video content. Such a procedure is able to reveal the semantic information contained in a video and is thus called video semantic recounting.
Figure~\ref{fig_recounting} shows the recounting result on videos from some exemplary events in MED and CCV, in which the top $5$ ranked concepts generated by our method are selected as concepts for each video. As can be seen, these concepts reveal the semantics contained in the videos, which verifies the effectiveness of the our discovered concepts in representing video semantics.


\subsection{Computational Cost}
The proposed Concept Bank is efficient computationally.
We calculate the running time on a 2.8GHz Windows workstation.
In average, for a event query, semantic matching needs $78$ seconds with Python.
For a video, the average time to generate a $100$ dimension concept score (including feature extraction) is around $110$ seconds with MATLAB.

\newpage

\newpage
\section{Conclusions and Future Work}
\label{sec_conclusion}
We have introduced Concept Bank, a concept library specifically designed for complex event representation in videos.
The library consists of $4,876$ concepts organized in a five-layer tree structure, in which the higher-level nodes correspond to the event hierarchies and the leaf nodes are the event-specific concepts.
To include as many real-world events in the library as possible, we collect all visually detectable events in WikiHow.
For each event, we use its article names in WikiHow to search Flickr and then discover relevant concepts from the tags associated with the crawled images of this event.
A Multiple Kennel Linear SVM classifier is then trained for each concept with the crawled images as the atomic concept detector.
In addition, we also develop a semantic matching method to determine relevant concepts for a coming event query.
Experiments over two event detection tasks verify the effectiveness of the proposed Concept Bank.
For future work, we will further extend the Concept Bank with motion concepts, which are trained based on the spatial-temporal features extracted from videos.

\newpage
{
\bibliographystyle{plain}
\bibliography{paper.bib}

\begin{thebibliography}{10}

\bibitem{MED}
\url{http://www.nist.gov/itl/iad/mig/med13.cfm}.

\bibitem{auto_attributes}
Tamara~L Berg, Alexander~C Berg, and Jonathan Shih.
\newblock Automatic attribute discovery and characterization from noisy web
  data.
\newblock In {\em ECCV}. 2010.

\bibitem{nltk}
Steven Bird.
\newblock Nltk: the natural language toolkit.
\newblock In {\em COLING/ACL on Interactive presentation sessions}, 2006.

\bibitem{FlickrConcept}
Jiawei Chen, Yin Cui, Guangnan Ye, Dong Liu, , and Shih-Fu Chang.
\newblock Event-driven semantic concept discovery by exploiting weakly-tagged
  internet images.
\newblock In {\em ICMR}, 2014.

\bibitem{imagenet}
Jia Deng, Wei Dong, Richard Socher, Li-Jia Li, Kai Li, and Li~Fei-Fei.
\newblock Imagenet: A large-scale hierarchical image database.
\newblock In {\em CVPR}, 2009.

\bibitem{duan2009domain}
Lixin Duan, Ivor~W Tsang, Dong Xu, and Stephen~J Maybank.
\newblock Domain transfer svm for video concept detection.
\newblock In {\em CVPR}, 2009.

\bibitem{liblinear}
Rong-En Fan, Kai-Wei Chang, Cho-Jui Hsieh, Xiang-Rui Wang, and Chih-Jen Lin.
\newblock Liblinear: A library for large linear classification.
\newblock {\em JMLR}, 2008.

\bibitem{gabor}
David~J Field et~al.
\newblock Relations between the statistics of natural images and the response
  properties of cortical cells.
\newblock {\em J. Opt. Soc. Am. A}, 1987.

\bibitem{ccv}
Yu-Gang Jiang, Guangnan Ye, Shih-Fu Chang, Daniel Ellis, and Alexander~C Loui.
\newblock Consumer video understanding: A benchmark database and an evaluation
  of human and machine performance.
\newblock In {\em ICMR}, 2011.

\bibitem{spm}
Svetlana Lazebnik, Cordelia Schmid, and Jean Ponce.
\newblock Beyond bags of features: Spatial pyramid matching for recognizing
  natural scene categories.
\newblock In {\em CVPR}, 2006.

\bibitem{rank_fusion}
Joon~Ho Lee.
\newblock Analyses of multiple evidence combination.
\newblock In {\em ACM SIGIR}, 1997.

\bibitem{object_bank}
Li-Jia Li, Hao Su, Li~Fei-Fei, and Eric~P Xing.
\newblock Object bank: A high-level image representation for scene
  classification \& semantic feature sparsification.
\newblock In {\em NIPS}, 2010.

\bibitem{conceptnet}
Hugo Liu and Push Singh.
\newblock ConceptnetÑa practical commonsense reasoning tool-kit.
\newblock {\em BT technology journal}, 2004.

\bibitem{liu_concept}
Jingen Liu, Qian Yu, Omar Javed, Saad Ali, Amir Tamrakar, Ajay Divakaran, Hui
  Cheng, and Harpreet~S Sawhney.
\newblock Video event recognition using concept attributes.
\newblock In {\em WACV}, 2013.

\bibitem{sift}
David~G Lowe.
\newblock Distinctive image features from scale-invariant keypoints.
\newblock {\em IJCV}, 2004.

\bibitem{LSCOM}
Milind Naphade, John~R Smith, Jelena Tesic, Shih-Fu Chang, Winston Hsu, Lyndon
  Kennedy, Alexander Hauptmann, and Jon Curtis.
\newblock Large-scale concept ontology for multimedia.
\newblock {\em Multimedia, IEEE}, 2006.

\bibitem{natarajan2012fusion}
Pradeep Natarajan, Shuang Wu, Shiv Vitaladevuni, Xiaodan Zhuang, Stavros
  Tsakalidis, Unsang Park, and Rohit Prasad.
\newblock Multimodal feature fusion for robust event detection in web videos.
\newblock In {\em CVPR}, 2012.

\bibitem{lbp}
Timo Ojala, Matti Pietikainen, and Topi Maenpaa.
\newblock Multiresolution gray-scale and rotation invariant texture
  classification with local binary patterns.
\newblock {\em TPAMI}, 2002.

\bibitem{gist}
Aude Oliva and Antonio Torralba.
\newblock Modeling the shape of the scene: A holistic representation of the
  spatial envelope.
\newblock {\em IJCV}, 2001.

\bibitem{kde}
Emanuel Parzen.
\newblock On estimation of a probability density function and mode.
\newblock {\em The annals of mathematical statistics}, 1962.

\bibitem{labelme}
Bryan~C Russell, Antonio Torralba, Kevin~P Murphy, and William~T Freeman.
\newblock Labelme: a database and web-based tool for image annotation.
\newblock {\em IJCV}, 2008.

\bibitem{action_bank}
Sreemanananth Sadanand and Jason~J Corso.
\newblock Action bank: A high-level representation of activity in video.
\newblock In {\em CVPR}, 2012.

\bibitem{tang2012latent}
Kevin Tang, Li~Fei-Fei, and Daphne Koller.
\newblock Learning latent temporal structure for complex event detection.
\newblock In {\em CVPR}, 2012.

\bibitem{classemes}
Lorenzo Torresani, Martin Szummer, and Andrew Fitzgibbon.
\newblock Efficient object category recognition using classemes.
\newblock In {\em ECCV}. 2010.

\bibitem{color}
Koen~EA Van De~Sande, Theo Gevers, and Cees~GM Snoek.
\newblock Evaluating color descriptors for object and scene recognition.
\newblock {\em TPAMI}, 2010.

\bibitem{soft_quantization}
Jan~C van Gemert, Cor~J Veenman, Arnold~WM Smeulders, and J-M Geusebroek.
\newblock Visual word ambiguity.
\newblock {\em TPAMI}, 2010.

\bibitem{wikihow}
By~Wikipedians.
\newblock {\em All About wikiHow}.
\newblock PediaPress.

\bibitem{yang2012deepnet}
Yang Yang and Mubarak Shah.
\newblock Complex events detection using data-driven concepts.
\newblock In {\em ECCV}. 2012.

\bibitem{bimodal}
Guangnan Ye, I-hong Jhuo, Dong Liu, Yugang Jiang, D.~T. Lee, and Shih-Fu Chang.
\newblock Joint audio-visual bi-modal codewords for video event detection.
\newblock In {\em ICMR}, 2012.

\end{thebibliography}
}

\end{document}